\documentstyle[12pt,epsfig]{article}
\begin{document}
\begin{flushright}
PPREPRINT IHEP-99-03\\
hep-ph/9902xxx
\end{flushright}

\begin{center}

{High twist contribution to the longitudinal structure function 
$F_L$ at high $x$}

\vspace{0.1in}

{\bf S.I. Alekhin }

\vspace{0.1in}
{\baselineskip=14pt
Institute for High Energy Physics, 142281 Protvino, Russia}

\begin{abstract}
We performed NLO QCD fit of 
combined SLAC/BCDMS/NMC DIS data at high $x$.
Model independent x-shape of 
high twist contribution to structure function $F_L$
is extracted. Twist-4 contribution is
found to be in fair agreement with the predictions of
infrared renormalon model. Twist-6 contribution
exhibits weak tendency to negative values, although
in the whole is compatible with zero within errors.
\end{abstract}
\end{center}

{\bf 1.} It is well known that on the basis of the  operator product expansion
the deep inelastic scattering (DIS) cross sections can be split 
into the leading
twist (LT) and the higher twist (HT) 
contributions. To the moment the LT 
contribution is fairly well understood
both from the theoretical and experimental points of view.
The HT contribution is not so well explored. 
The theoretical investigations of HT 
meet with the difficulties because
in the region where HT are most important the perturbative QCD calculations
cannot be applied.
There are only semi-qualitative 
phenomenological models for HT calculation
in literature. These models are often based on the 
phenomenological considerations and contain adjusted 
parameters, which are to  be determined from the
experimental data.
Unfortunately, relevant experimental 
data, especially on the longitudinal structure function $F_L$,
are sparse, come from different experiments and 
therefore are difficult for interpretation.
There are
estimations of twist-4 contribution to the structure function $F_2$
\cite{VIR,KOT,AL} obtained from the combined fit to SLAC/BCDMS 
and SLAC/BCDMS/NMC data 
\cite{BCDMS,SLAC,NMC}. Extraction of HT contribution 
to $F_3$  was obtained in \cite{KAT,AK} from the fit to CCFR data \cite{CCFR}.
These estimations are model independent, i.e. do not imply
any x-dependence of HT, and hence 
a phenomenological formula
can be easily fitted to them.
As to the experimental data on HT contribution to $F_L$, 
to the moment they are available only from 
the QCD motivated fits to world data on 
the structure function $R=\sigma_L/\sigma_T$.
The first fit of this kind was presented in \cite{R1990}.
This fit was recently renewed in \cite{E143} with the inclusion of new data
from the experiments E-143 and E-140X.
The world data on $R$ were also analyzed using a QCD based model
with the account of HT contribution \cite{BODEK}.  
As far some models predict the HT contribution 
to the structure function $F_L$, the comparison of these
models with the data requires extraction of the HT contribution to $F_L$
from the data on the HT contribution to $R$ and $F_2$. This causes 
the problems with interpolation between data points and  
error propagation.
In addition, the HT contributions to 
$R$, obtained in all these fits  
are model dependent, i.e. a priori suppose 
a certain x-dependence of HT.

{\bf 2.} In this paper we present the results of the analysis of
DIS data aiming to obtain the  estimation of the 
model independent HT contribution to $F_L$. The work is the 
continuation of our previous analysis \cite{AL}, where 
estimation of the HT contribution to $F_2$ was obtained.
Our consideration is limited by the region of 
$x>0.3$, where the non-singlet approximation is valid.
A data cut $x<0.75$
was made to leave the region where deuteron effects are small. 
At the beginning 
the ansatz used in this work is essentially the same as in \cite{AL}.
We fitted the data within NLO QCD with the 
inclusion of 
target mass correction (TMC) \cite{TMC}
and twist-4 contribution 
in factorized form:
\begin{displaymath}  
F_2^{(P,D),HT}(x,Q)=F_2^{(P,D),TMC}(x,Q)
\Bigl[1+\frac{h_2^{(P,D)}(x)}{Q^2}\Bigr],
\end{displaymath}  
\begin{displaymath}
F_2^{(P,D),TMC}(x,Q)=\frac{x^2}{\tau^{3/2}}\frac{F_2^{(P,D),LT}(\xi,Q)}{\xi^2}
+6\frac{M^2}{Q^2}\frac{x^3}{\tau^2}\int^{1}_{\xi}dz
\frac{F_2^{(P,D),LT}(z,Q)}{z^2},
\end{displaymath}
where $F_2^{(P,D),LT}(x,Q)$ are the LT terms, 
\begin{displaymath}
\xi=\frac{2x}{1+\sqrt{\tau}},~~~~\tau=1+\frac{4M^2x^2}{Q^2},
\end{displaymath}
$M$ is nucleon mass, $x$ and $Q^2$ are regular lepton scattering variables.
The LT 
structure functions of 
proton and neutron were
parametrized at the initial
value of $Q_0^2=9 GeV^2$ as follows:
\begin{displaymath}  
F^p_2(x,Q_0)=A_{p}x^{a_{p}}(1-x)^{b_{p}}\frac{2}{N_p}
\end{displaymath}  
\begin{displaymath}  
F^n_2(x,Q_0)=A_{n}x^{a_{n}}(1-x)^{b_{n}}\frac{1}{N_n}.
\end{displaymath}  
Here conventional normalization factors $N_p$ and $N_n$ are
\begin{displaymath}  
N_{p,n}=\int_0^1dxx^{a_{p,n}-1}(1-x)^{b_{p,n}}.
\end{displaymath}  
These distributions were evolved 
in NLO QCD approximation 
within $\overline{MS}$ factorization scheme.
The functions $h_2^{(P,D)}(x)$ were parametrized in the model independent 
way: their values at $x=0.3,0.4,0,5,0.6,0.7,0.8$ were fitted,
between these points the functions were linearly interpolated.

Comparing to the work \cite{AL} we added to the analysis the
NMC data \cite{NMC} 
(30 points on proton and 30 points on deuterium targets) 
The number of data for each experiment
and target are presented in Table I. We accounted for point-to-point
correlations of data due to systematic errors analogously to our 
previous papers \cite{PDF,AL,AK}. 
Systematic errors were convoluted into correlation 
matrix
\begin{displaymath}
C_{ij}=\delta_{ij}\sigma_i\sigma_j+f_if_j(\vec{s}_i^K \cdot \vec{s}_j^K),
\end{displaymath}
where vectors $\vec s_i^K$ contain systematic errors, index $K$ runs
through data subsets which are uncorrelated between each other, $i$ and $j$
run through data points of these data subsets. 
Minimized functional has the form  
\begin{displaymath}
\chi^2=\sum_{K,i,j}
(f_i/\xi_K-y_i)E_{ij}(f_j/\xi_K-y_j), 
\end{displaymath}
where $E_{i,j}$ is the inverse of $C_{i,j}$.
Dimension of $\vec s_i^K$ differs for various data set, the concrete
numbers for each experiment are present in Table I.
The factors $\xi_K$ were introduced to allow for renormalization 
of some data sets. In our analysis these factors were released for 
the old SLAC experiments in view of possible normalization errors 
discussed in \cite {WHITT}. As for E-140, BCDMS and NMC data subsets,
we fixed these factors at 1 and accounted for their
published normalization errors in the general correlation matrix.

\begin{table}
\begin{center}
\caption{The number of data points (NDP) and the number of independent
systematic errors (NSE)
for the analysed data sets.}
\begin{tabular}{cccc} 
Experiment&NDP(proton)&NDP(deuterium)&NSE\\  
BCDMS  &223&162 &9        \\  
E-49A  &47&47   &3   \\  
E-49B  &109&102 &3     \\  
E-61   &6&6     &3 \\  
E-87   &90&90   &3   \\  
E-89A   &66&59  &3    \\  
E-89B  &70&59   &3   \\  
E-139   &--&16  &3    \\  
E-140  &--&31   &4   \\  
NMC    &30&30 &13   \\   
TOTAL  &641&602 &47     \\
\end{tabular}
\end{center}
\end{table}

At the first stage of the analysis
we reduced all $F_2$ data including BCDMS and NMC
to the common value of $R$ \cite{R1990}.
The results of the fit within this ansatz are presented in Table II 
(column 2). For the comparison at column 1 we also presented the results 
of analogous fit from from \cite{AL}, performed without the NMC data. 
Addition of the NMC data increased the value 
of $\alpha_s(M_Z)$, but within one standard deviation.
In the whole the figures from column 2 are compatible 
with the results of earlier analysis \cite {AL}.

\begin{table}
\begin{center}
\caption{The results of the fits with factorized parametrization 
of HT. The parameters $\xi$ describe the renormalization
of the old SLAC data, $h^{P,D}_{2,(3,4,5,6,7,8)}$ are the fitted values of 
the HT contribution at $x=0.3,0.4,0.5,0.6,0.7,0.8$.
For the description of the columns see text.}
\begin{tabular}{ccc}     
                 &       1           &       2        \\ 
$A_p$            & $0.516\pm0.022$      & $0.514\pm0.021$   \\ 
$a_p$            & $0.765\pm0.028$   &  $0.766\pm0.028$   \\ 
$b_p$            & $3.692\pm0.032$   & $3.690\pm0.032$   \\ 
$A_n$            & $4.8\pm4.1$       & $4.8\pm3.9$       \\ 
$a_n$            & $0.118\pm0.097$   & $0.119\pm0.095$    \\ 
$b_n$            & $3.51\pm0.11$     & $3.51\pm0.11$     \\ 
$\alpha_s(M_Z)$  & $0.1180\pm0.0017$ & $0.1187\pm0.0016$ \\ 
$h^P_{2,3}$          & $-0.120\pm0.017$ & $-0.122\pm0.017$ \\ 
$h^P_{2,4}$          & $-0.046\pm0.025$ & $-0.054\pm0.025$ \\ 
$h^P_{2,5}$          & $0.059\pm0.043$  & $0.043\pm0.042$    \\ 
$h^P_{2,6}$          & $0.392\pm0.076$  & $0.363\pm0.074$  \\ 
$h^P_{2,7}$          & $0.82\pm0.13$    & $0.77\pm0.12$  \\ 
$h^P_{2,8}$          & $1.54\pm0.25$    & $1.47\pm0.24$  \\ 
$h^D_{2,3}$          & $-0.123\pm0.018$ & $-0.125\pm0.018$ \\ 
$h^D_{2,4}$          & $-0.003\pm0.026$ & $-0.012\pm0.025$ \\ 
$h^D_{2,5}$          & $0.162\pm0.043$  & $0.145\pm0.042$  \\ 
$h^D_{2,6}$          & $0.439\pm0.076$  & $0.410\pm0.073$  \\ 
$h^D_{2,7}$          & $0.79\pm0.12$    & $0.75\pm0.11$ \\ 
$h^D_{2,8}$          & $1.87\pm0.26$    & $1.81\pm0.25$ \\ 
$\xi_{P,49A}$    & $1.016\pm0.018$  & $1.016\pm0.016$   \\ 
$\xi_{D,49A}$    & $1.006\pm0.017$   & $1.008\pm0.015$   \\ 
$\xi_{P,49B}$    & $1.028\pm0.018$   & $1.028\pm0.015$   \\ 
$\xi_{D,49B}$    & $1.012\pm0.017$   & $1.013\pm0.015$   \\ 
$\xi_{P,61}$     & $1.021\pm0.021$   & $1.021\pm0.019$   \\ 
$\xi_{D,61}$     & $1.004\pm0.019$   & $1.006\pm0.018$   \\ 
$\xi_{P,87}$     & $1.025\pm0.017$   & $1.025\pm0.015$   \\ 
$\xi_{D,87}$     & $1.012\pm0.017$   & $1.013\pm0.015$   \\ 
$\xi_{P,89A}$    & $1.028\pm0.021$   & $1.029\pm0.019$   \\ 
$\xi_{D,89A}$    & $1.004\pm0.021$   & $1.006\pm0.019$   \\ 
$\xi_{P,89B}$    & $1.022\pm0.017$   & $1.021\pm0.015$   \\ 
$\xi_{D,89B}$    & $1.007\pm0.017$   & $1.008\pm0.015$   \\ 
$\xi_{D,139}$    & $1.009\pm0.017$   & $1.010\pm0.015$   \\ 
$\chi^2/NDP$     & $1178.9/1183$     & $1258.4/1243$       \\ 
\end{tabular}
\end{center}
\end{table}

The next step of our analysis was to change the form of HT contribution 
from factorized to an additive one: 
\begin{displaymath}  
F_2^{(P,D),HT}(x,Q)=F_2^{(P,D),TMC}(x,Q)+H_2^{(P,D)}(x)\frac{1~GeV^2}{Q^2}
\end{displaymath}  
with $H_2^{(P,D)}(x)$ parametrized in the model independent 
form analogously to $h_2^{(P,D)}(x)$.
We preferred to switch to this form because for the  factorized
parametrization HT term contains latent log factor
and twist-6 contribution originating from $F_2^{LT}(x,Q)$ and
target mass corrections respectively. In addition,
this form is more convenient for comparison with some model predictions.
The results of the fit with additive HT parametrization are presented 
in column 1 of Table III and Fig.1. 
For the comparison we also pictured in Fig.1 the value 
of $F_2^{TMC}(x,Q)\cdot h_2(x)$ 
for the fit from column 2 of Table II 
with factorized form of HT, calculated at $Q^2=2.5~GeV^2$.
One can see that the switching of the form leads to small decrease
of HT terms at high $x$.
Besides, their errors became smaller, meanwhile errors of the other 
parameters increased.
We connect this effect with that the additive HT form is not so constrained
as the factorized one. The change of the value of HT contribution is
amplified due to large correlations of the HT terms 
with other parameters, especially with $\alpha_s$ (see 
discussion below).

\begin{figure}
\centerline{\epsfig{file=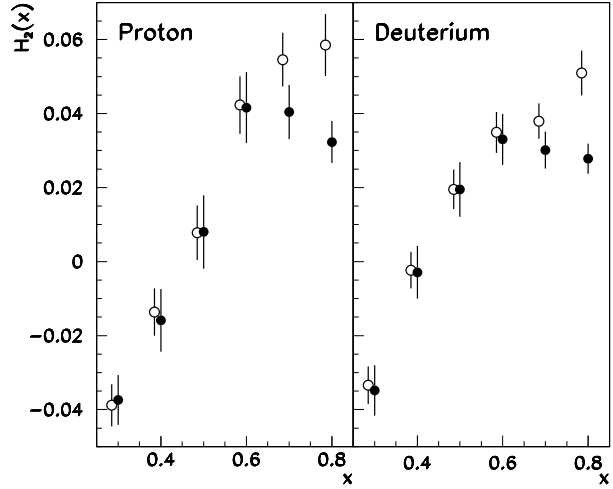,width=8cm}}
\caption{The results of fits with different forms
of HT contribution (full circles - additive, 
empty circles - factorized).
For the better view the points are shifted 
to left/right along x-axis.}
\end{figure}

To extract $F_L$ we replaced data on $F_2$ by the 
the data on differential cross sections
and fitted them using the formula
\begin{displaymath}
\frac{d^2\sigma}{dxdy}=\frac{4\pi\alpha^2(s-M^2)}{Q^4}
\biggl[\biggl(1-y-\frac{(Mxy)^2}{Q^2}\biggr)F_2^{HT}(x,Q)+
\end{displaymath}
\begin{displaymath}
+\biggl(1-2\frac{m_l^2}{Q^2}\biggr)
\frac{y^2}{2}\biggl(F_2^{HT}(x,Q)-F_L^{HT}(x,Q)\biggr)\biggl],
\end{displaymath}
where $s$ is total c.m.s. energy, $m_l$ is 
scattered lepton mass and 
$y$ - lepton scattering variable.
With the test purposes we first  performed the
fit with $F_L$ form motivated by $R_{1990}$ parametrization \cite{R1990}:
\begin{displaymath}
F_L^{(P,D),HT}(x,Q)=F_2^{(P,D),HT}(x,Q)\biggl[1-
\frac{1+4M^2x^2/Q^2}{1+R(x,Q)}\biggr],
\end{displaymath}
\begin{displaymath}
R(x,Q)=\frac{b_1}{2\ln(Q/0.2)}
\biggl[1+\frac{12Q^2}{Q^2+1}\cdot\frac{0.125^2}{0.125^2+x^2}\biggr]
+b_2\frac{1~GeV^2}{Q^2}+b_3\frac{1~GeV^4}{Q^4+0.3^2}
\end{displaymath}
with fitted parameters $b_{1,2,3}$ and $Q$ measured in $GeV$.
This parametrization is constructed from the 
term with log $Q$-behaviour, 
which mimic LT contribution and  
HT terms with power  $Q$-behaviour.
The resulting values of 
the parameters $b_1=0.100\pm0.042$, $b_3=0.46\pm0.11$, 
$b_3=-0.14\pm0.18$
are in agreement with the values obtained in \cite{R1990}
($b_1=0.064$, $b_3=0.57$, $b_3=-0.35$).
In our fit twist-6 contribution to $R(x,Q)$ is
compatible with zero. 
We can note in this connection, that 
the correlations between the parameters is large
(see Table IV) and as a consequence
the data used in the fit have limited potential both in 
separation of twist-4/twist-6 
and log/power contributions to $R$.

\begin{table}
\begin{center}
\caption{The results of the fits with additive parametrization 
of HT. Figures in parenthesis are global correlation coefficients.
The parameters $\xi$ describe the renormalization
of the old SLAC data, $H^{P,D}_{2,(3,4,5,6,7,8)}$
and $H^{P,D}_{L,(3,4,5,6,7,8)}$ are the fitted values of 
the HT contribution at $x=0.3,0.4,0.5,0.6,0.7,0.8$.
For the description of the columns see text.}
\small
\begin{tabular}{cccc}      
                 &       1           &       2         &       3 \\ 
$A_p$            & $0.478\pm0.021$   & $0.485\pm0.023$   & $0.486\pm0.022$ \\ 
$a_p$            & $0.830\pm0.034$   & $0.816\pm0.037$   & $0.817\pm0.035$  \\ 
$b_p$            & $3.798\pm0.043$   & $3.791\pm0.047$   & $3.792\pm0.045$   \\ 
$A_n$            & $4.8\pm4.5$       & $4.9\pm4.8$       & $4.7\pm4.7$   \\ 
$a_n$            & $0.12\pm0.11$   & $0.12\pm0.13$     & $0.12\pm0.12$  \\ 
$b_n$            & $3.58\pm0.14$     & $3.59\pm0.15$      & $3.58\pm0.14$  \\ 
$\alpha_s(M_Z)$  & $0.1190\pm0.0021$ & $0.1170\pm0.0021$ & $0.1170\pm0.0021$ \\ 
$H^P_{2,3}$          & $-0.0374\pm0.0067$  & $-0.0303\pm0.0068$  & $-0.0273\pm0.0067$  \\ 
$H^P_{2,4}$          & $-0.0159\pm0.0085$  & $-0.0031\pm0.0084$  & $-0.0059\pm0.0082$  \\ 
$H^P_{2,5}$          & $0.0080\pm0.0099$   & $0.0176\pm0.0096$     & $0.0189\pm0.0095$  \\ 
$H^P_{2,6}$          & $0.0416\pm0.0096$   & $0.0497\pm0.0095$   & $0.0494\pm0.0092$  \\ 
$H^P_{2,7}$          & $0.0404\pm0.0073$    & $0.0529\pm0.0077$     & $0.0501\pm0.0073$  \\ 
$H^P_{2,8}$          & $0.0323\pm0.0057$     & $0.0376\pm0.0081$     & $0.0400\pm0.0070$  \\ 
$H^D_{2,3}$          & $-0.0348\pm0.0068$  & $-0.0205\pm0.0066$  & $-0.0221\pm0.0065$  \\ 
$H^D_{2,4}$          & $-0.0029\pm0.0071$  & $0.0057\pm0.0069$  & $0.0067\pm0.0068$  \\ 
$H^D_{2,5}$          & $0.0195\pm0.0074$   & $0.0274\pm0.0071$   & $0.0269\pm0.0070$   \\ 
$H^D_{2,6}$          & $0.0330\pm0.0069$   & $0.0381\pm0.0068$   & $0.0380\pm0.0067$   \\ 
$H^D_{2,7}$          & $0.0301\pm0.0050$     & $0.0363\pm0.0051$     & $0.0372\pm0.0050$   \\ 
$H^D_{2,8}$          & $0.0278\pm0.0041$     & $0.0354\pm0.0061$     & $0.0341\pm0.0053$  \\ 
$H^P_{L,3}$          & --  & $0.045\pm0.027$  & -- \\ 
$H^P_{L,4}$          & --  & $0.036\pm0.024$  & --\\ 
$H^P_{L,5}$          & --   & $-0.009\pm0.023$     & --\\ 
$H^P_{L,6}$          & --   & $-0.012\pm0.017$   & --\\ 
$H^P_{L,7}$          & --    & $0.020\pm0.013$     & --\\ 
$H^P_{L,8}$          & --     & $0.004\pm0.019$     & --\\ 
$H^D_{L,3}$          & --  & $0.106\pm0.016$  & $0.095\pm0.014$  \\ 
$H^D_{L,4}$          & --  & $0.044\pm0.013$  & $0.049\pm0.012$  \\ 
$H^D_{L,5}$          & --   & $0.0343\pm0.0089$   & $0.031\pm0.0087$  \\ 
$H^D_{L,6}$          & --   & $0.009\pm0.010$   & $0.0068\pm0.0094$  \\ 
$H^D_{L,7}$          & --     & $0.0161\pm0.0078$     & $0.0195\pm0.0069$  \\ 
$H^D_{L,8}$          & --     & $0.021\pm0.018$     & $0.016\pm0.013$  \\ 
$\xi_{P,49A}$    & $1.013\pm0.016$   & $1.017\pm0.016$   & $1.017\pm0.016$  \\ 
$\xi_{D,49A}$    & $1.005\pm0.015$   & $1.007\pm0.015$   & $1.009\pm0.015$  \\ 
$\xi_{P,49B}$    & $1.023\pm0.015$   & $1.039\pm0.016$   & $1.031\pm0.015$  \\ 
$\xi_{D,49B}$    & $1.010\pm0.015$   & $1.011\pm0.015$   & $1.014\pm0.015$  \\ 
$\xi_{P,61}$     & $1.017\pm0.019$   & $1.026\pm0.019$   & $1.028\pm0.019$  \\ 
$\xi_{D,61}$     & $1.004\pm0.018$   & $1.018\pm0.018$   & $1.018\pm0.018$  \\ 
$\xi_{P,87}$     & $1.019\pm0.015$   & $1.029\pm0.016$   & $1.025\pm0.015$  \\ 
$\xi_{D,87}$     & $1.008\pm0.015$   & $1.007\pm0.015$   & $1.011\pm0.015$  \\ 
$\xi_{P,89A}$    & $1.029\pm0.019$   & $1.057\pm0.023$   & $1.041\pm0.020$  \\ 
$\xi_{D,89A}$    & $1.005\pm0.019$   & $1.011\pm0.020$   & $1.011\pm0.020$  \\ 
$\xi_{P,89B}$    & $1.016\pm0.015$   & $1.024\pm0.016$   & $1.021\pm0.015$  \\ 
$\xi_{D,89B}$    & $1.003\pm0.015$   & $1.004\pm0.015$   & $1.007\pm0.015$  \\ 
$\xi_{D,139}$    & $1.010\pm0.015$   & $1.012\pm0.015$   & $1.014\pm0.015$  \\ 
$\chi^2/NDP$     & $1274.3/1223$       & $1248.0/1243$       & $1255.6/1243$   \\ 
\end{tabular}
\end{center}
\normalsize
\end{table}
\begin{table}
\begin{center}
\caption{Correlation coefficients for the parameters of the  
fit motivated by $R_{1990}$ parametrization.}
\begin{tabular}{cccc}      
                 &       $b_1$           &       $b_2$         &      $b_3$         \\ 
$b_1$            & 1.   & -0.61   & 0.38   \\ 
$b_2$            & -0.61   & 1.   & -0.87   \\ 
$b_3$            & 0.38   & -0.87   & 1.   \\ 
\end{tabular}
\end{center}
\end{table}

\begin{figure}
\centerline{\epsfig{file=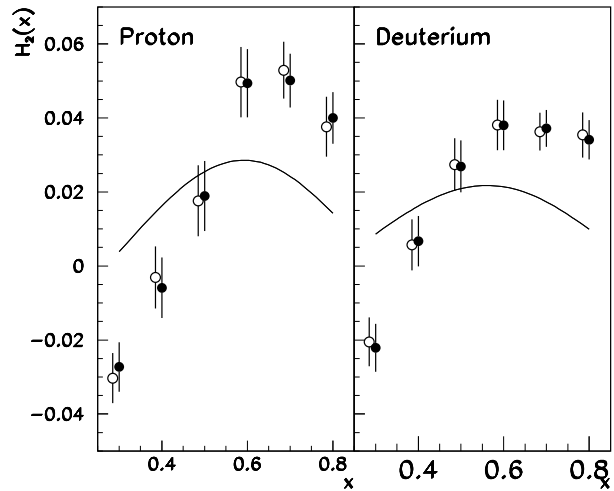,width=8cm}}
\caption{The values of $H_2(x)$ from the fits with model independent form of 
HT contribution to $F_L$ (empty circles -- unconstrainted fit, 
full circles -- the fit with the constraint $H_L^D(x)=H_L^P(x)$
For the better view the points are shifted 
to left/right along x-axis. The curves are predictions of IRR model
for $Q^2=2~GeV^2$.}
\end{figure}

\begin{figure}
\centerline{\epsfig{file=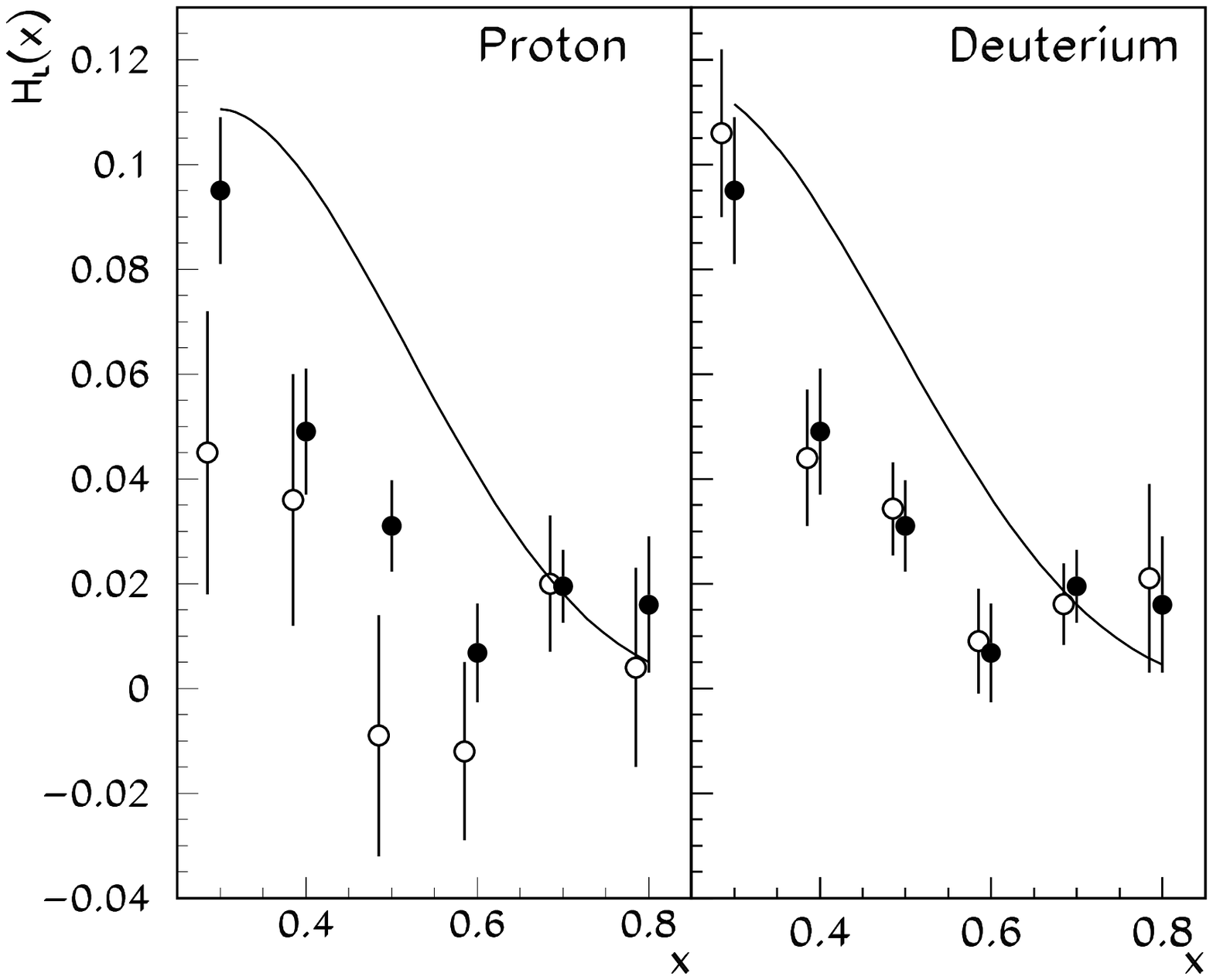,width=8cm}}
\caption{The same as Fig.2 for $H_L(x)$.}
\end{figure}

To achieve more precise determination of HT contribution to $F_L$
one can substitute for the LT contribution a perturbative 
QCD formula \cite{AM} instead of phenomenological 
log term. In the leading order on $\alpha_s$ and for nonsinglet case
it looks like
\begin{displaymath}
F_L^{(P,D),LT}(x,Q)=
\frac{\alpha_s(Q)}{2\pi}\frac{8}{3}x^2
\int^{1}_{x}\frac{dz}{z^3}F_2^{(P,D),LT}(z,Q).
\end{displaymath}
We performed the fit using the QCD expression for $F_L$ 
with the account of TMC
\begin{displaymath}
F_L^{(P,D),TMC}(x,Q)=F_L^{(P,D),LT}(x,Q)+\frac{x^2}{\tau^{3/2}}
\frac{F_2^{(P,D),LT}(\xi,Q)}{\xi^2}(1-\tau)+
\end{displaymath}
\begin{displaymath}
+\frac{M^2}{Q^2}\frac{x^3}{\tau^2}(6-2\tau)
\int^{1}_{\xi}dz\frac{F_2^{(P,D),LT}(z,Q)}{z^2}
\end{displaymath}
and additive form of HT contribution to $F_L$, i.e.
\begin{displaymath}
F_L^{(P,D),HT}(x,Q)=F_L^{(P,D),TMC}(x,Q)+H_L^{(P,D)}(x)\frac{1~GeV^2}{Q^2},
\end{displaymath}
where $H_L^{(P,D)}(x)$ is parametrized in the model independent form
analogously to $H_2^{(P,D)}(x)$ and $h_2^{(P,D)}(x)$.
The results of this fit are presented in column 2 of Table III
and in Figs.2,3. One can see that with model independent parametrization
of HT contribution to $F_L$ 
the renormalization factors for old SLAC data 
noticeably increased. We remind in this connection that 
in the source paper \cite{WHITT} these data were renormalized 
using the data of dedicated SLAC-E-140 experiment. As far
the latest did not reported data on proton target, 
renormalization of proton data was performed using bridging
through SLAC-E-49B data. There is no possibility 
to conclusively choose between our renormalization 
scheme and the one used in \cite {WHITT}. 
More proton data are necessary to adjust 
old SLAC results. As to deuteron data, their normalization factors
practically do not deviate from 1. The errors of $H^D_L(x)$,
due to SLAC-E-140 deuteron data,
are significantly smaller, than for $H^P_L(x)$. 
In view of large errors of the last,
HT contributions to $F_L$ for proton and deuteron 
are compatible within the errors and we performed
one more fit imposing the constraint $H^P_L(x)=H^D_L(x)$.
The results of this fit are presented in column 3 of Table IV
and in Figs. 2,3. One can see that $\chi^2$ obtained in this fit 
is practically equal to the value
of $\chi^2$ obtained in the fit 
without constraints minus the number of 
additional parameters. 
Comparison with Fig.1 shows, that if 
$F_L$ is fitted, the HT contribution to $F_2$ slightly growth..
The value of strong coupling constant is insensitive 
to the constraint imposing. Their value
\begin{displaymath}  
\alpha_s(M_Z)=0.1170\pm0.0021(stat+syst),
\end{displaymath}  
correspond to 
\begin{displaymath}  
\Lambda^{(3)}_{\overline{MS}}=337\pm29(stat+syst)~MeV,
\end{displaymath}  
obtained at $Q^2=2~GeV^2$ with the help of relation
\begin{displaymath}  
\alpha_s(Q)=\frac{2\pi}{\beta_0\ln(Q/\Lambda)}
\Bigl[1-\frac{2\pi}{\beta_0\beta}
\frac{\ln(2\ln(Q/\Lambda)}
{\ln(Q/\Lambda)}\Bigr], 
\end{displaymath}  
where
\begin{displaymath}
\beta_0=11-\frac{2}{3}n_f,~~~~\beta=\frac{2\pi\beta_0}{51-\frac{19}{3}n_f}
\end{displaymath}
and $n_f=3$.

\begin{figure}
\centerline{\epsfig{file=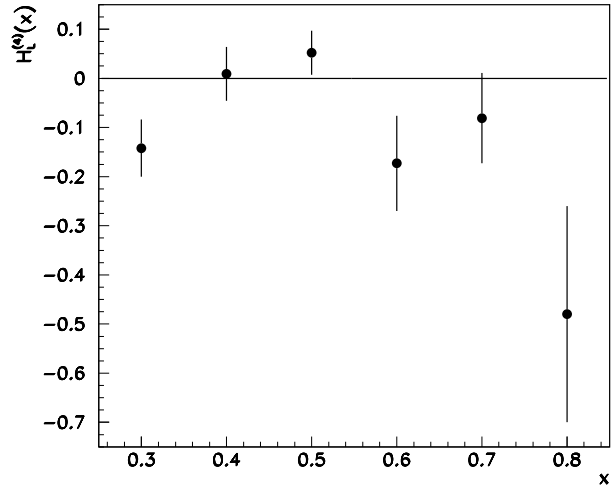,width=8cm}}
\caption{The dependence of twist-6 contribution to $F_L$ on $x$.}
\end{figure}

We checked how are the analysed data sensitive to the 
twist-6 contribution to $F_L$. For this purpose we 
fitted data with $F_L(x,Q)$ expressed as
\begin{displaymath}
F_L^{(P,D),HT}(x,Q)=F_L^{(P,D),TMC}(x,Q)+H_L(x)\frac{1~GeV^2}{Q^2}
+H_L^{(4)}(x)\frac{1~GeV^4}{Q^4},
\end{displaymath}
where functions $H_L^{(4)}(x)$ and $H_L(x)$ were the same for 
proton and deuteron 
and parametrized in model independent way.
The resulting behaviour of $H_L^{(4)}(x)$ is presented in Fig. 4.
One can observe the trend to the negative values at highest $x$, but with 
poor statistical significance. The $\chi^2$ decrease in this fit 
is about 25 (remind that standard deviation of $\chi^2$ is $\sqrt{2\cdot NDF}$, 
i.e. is about 40 in our analysis).
These results are in correspondence with the estimates of 
twist-6 contribution to $R(x,Q)$ presented above - 
the fitted twist-6 contribution to $F_L$ is at the level of 
one standard deviation. 
In our studies we did not completely accounted 
TMC correction of the order of $O(M^4/Q^4)$ and then, 
in a rigorous treatment,
$H_L^{(4)}$ does not correspond to a pure dynamical 
twist-6 effects. Meanwhile the contribution of the omitted terms 
to $H_L^{(4)}$ estimated using formula from \cite{TMC},
do not exceed 0.001
in our region of $x$ and hence can be neglected.

\begin{figure}
\centerline{\epsfig{file=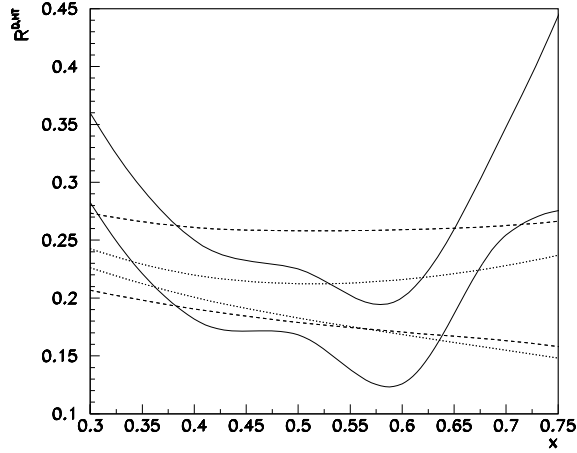,width=8cm}}
\caption{One standard deviation bands for our value of $R^{D,HT}(x,Q)$ (full 
lines), $R_{1990}$ (dashed lines) and $R_{1998}$ (dotted lines) at 
$Q^2=2~GeV^2$ Our value of $R$ is obtained as spline interpolation between 
points $x=0.3,0.4,0.5,0.6,0.7,0.8$.} 
\end{figure}

\begin{figure}
\centerline{\epsfig{file=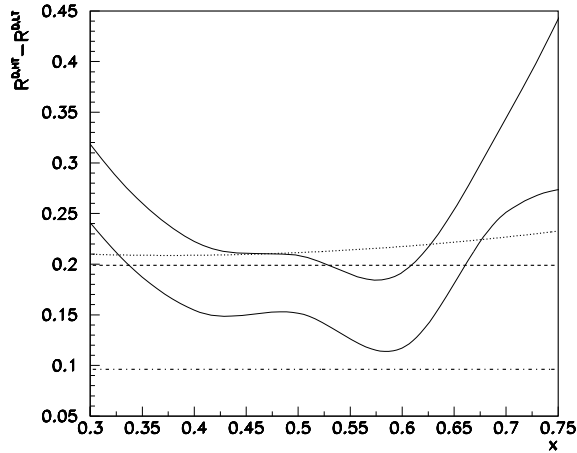,width=8cm}}
\caption{One standard deviation bands
for HT+TMC contribution to $R$ (full lines). 
Dashed and dotted lines correspond to power contributions
to $R_{1990}$ and $R_{1998}$. Dashed-dotted 
line represents the power term of BRY parametrization [14].}
\end{figure}

\begin{figure}
\centerline{\epsfig{file=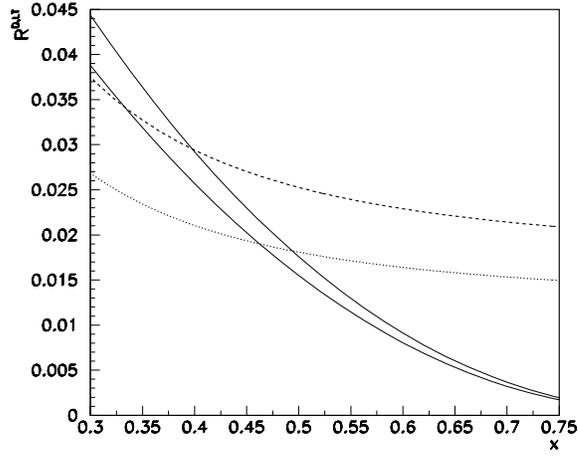,width=8cm}}
\caption{One standard deviation bands
for LT contribution to $R$ (full lines). 
Dashed and dotted lines correspond to log contributions
to $R_{1990}$ and $R_{1998}$.}
\end{figure}

\begin{figure}
\centerline{\epsfig{file=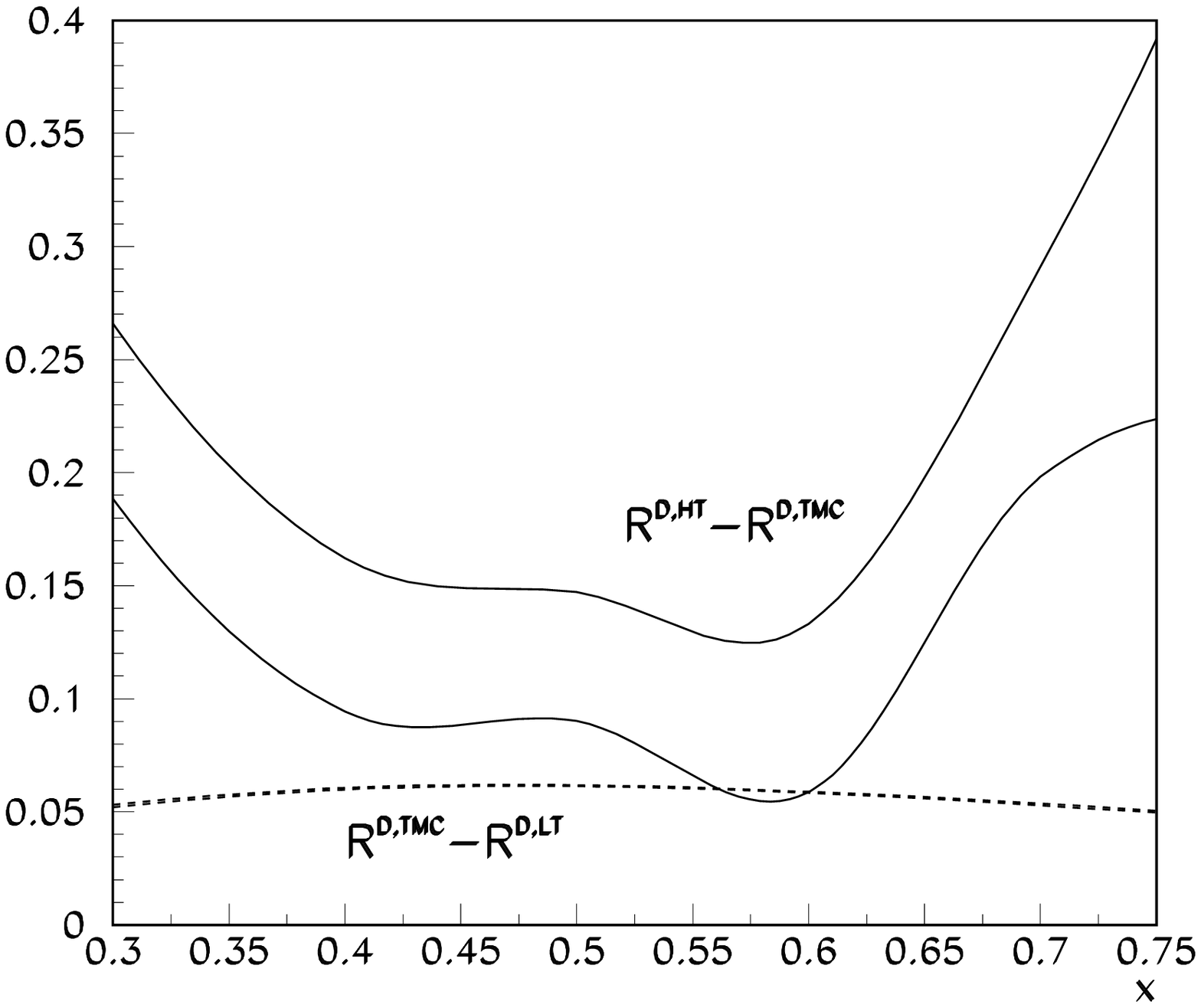,width=8cm}}
\caption{One standard deviation bands
for HT contribution (full lines) and TMC contribution (dashed lines)
to $R$. Lower and upper bands for TMC are practically indistinguishable.}
\end{figure}

{\bf 3.} For the comparison with other parametrizations 
we calculated deuteron $R(x,Q)$
using the relation
\begin{displaymath}
R^{D,HT}(x,Q)=\frac{F_2^{D,HT}(x,Q)}{F_2^{D,HT}(x,Q)-F_L^{D,HT}(x,Q)}
\biggl[1+\frac{4M^2x^2}{Q^2}\biggr]-1
\end{displaymath}
and the parameters values from column 3 of Table III.
The obtained values of $R$ are presented in Fig. 5 together with 
$R_{1990}$ \cite{R1990} and $R_{1998}$ \cite{E143} parametrizations. 
In average all three parametrizations coincide within errors, 
although our curves lay 
higher at the edges of x-region and lower 
in the middle.  

The same tendency is valid for HT+TMC contribution 
to $R$, evaluated as the difference between 
$R^{D,HT}(x,Q)$ and $R^{D,LT}(x,Q)$, where   
\begin{displaymath}
R^{D,LT}(x,Q)=\frac{F_L^{D,LT}(x,Q)}{F_2^{D,LT}(x,Q)-F_L^{D,LT}(x,Q)}.
\end{displaymath}
This contribution is presented in Fig.6 together with the power
parts of the variant b) of $R_{1990}$ and $R_{1998}$ parametrizations.
Since error bands are not available for latest ones, only central 
values are pictured. Relative difference between log parts of 
$R_{1990}$ and $R_{1998}$ parametrizations 
and $R^{D,LT}(x,Q)$ is very large at highest $x$, although absolute
difference is not significant due to smallness of these terms.
In Fig.8 we present  
the TMC contribution to $R$, calculated as $R^{D,TMC}- R^{D,LT}$,
where
\begin{displaymath}
R^{D,TMC}(x,Q)=\frac{F_2^{D,TMC}(x,Q)}{F_2^{D,TMC}(x,Q)
-F_L^{D,TMC}(x,Q)}
\biggl[1+\frac{4M^2x^2}{Q^2}\biggr]-1,
\end{displaymath}
and the HT contribution, calculated as $R^{D,HT}- R^{D,TMC}$.
It is seen that the 
TMC contribution also is significantly smaller at large  
$x$ than the HT one, so the latest is dominating 
in this region. Meanwhile we should note in this connection 
that as it was observed in \cite{KAT,YB}, account of NNLO QCD 
can diminish the value of HT contribution.
This effect can be seen in Fig.6, where we 
present power part of $R$ parametrization \cite{BODEK}, obtained with 
the partial account of NNLO. 

\begin{figure}
\centerline{\epsfig{file=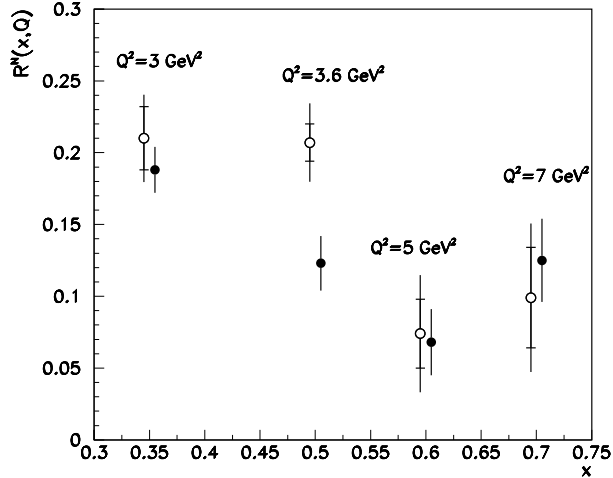,width=8cm}}
\caption{The E140X data on nucleon $R(x,Q)$ (empty circles).
Inner bars correspond to statistic errors, total bars -- 
statistic and systematic  combined in quadratures.
Also are presented calculations of deuteron $R(x,Q)$ performed
on our fit (full circles). For the better view the points are shifted 
to left/right along x-axis.}
\end{figure}

\begin{table}
\begin{center}
\caption{Coefficients of correlations between the HT parameters 
and $\alpha_s(M_Z)$. Figures in parenthesis are global correlation 
coefficients for the HT parameters.}
\begin{tabular}{cccc}      
   x           & $H^P_2(x)$ & $H^D_2(x)$ & $H_L(x)$  \\ 
0.3            & -0.507(0.914)& -0.707(0.964)  & -0.029(0.868) \\ 
0.4            & -0.847(0.955)& -0.910(0.976)  & 0.122(0.852)  \\ 
0.5            & -0.909(0.968)& -0.944(0.987)  & 0.244(0.870)  \\ 
0.6            & -0.905(0.971)& -0.917(0.977)  & 0.222(0.866)  \\ 
0.7            & -0.871(0.972)& -0.901(0.980)  & 0.072(0.894)  \\ 
0.8            & -0.460(0.868)& -0.429(0.875)  & 0.156(0.870)  \\ 
\end{tabular}
\end{center}
\end{table}

In Fig.9 data of SLAC-E140X experiment on nucleon $R(x,Q)$\cite{E140X},
which were not included in the fit, 
are compared with our $R^D(x,Q)$, calculated at the parameters values
from column 3 of Table III. One can observe a good agreement 
of the data and our parametrization.
We also compared the results of model independent analysis with 
a predictions of infrared renormalon model (IRR) \cite{IRR}.
In this model the HT x-dependence is connected with the 
LT x-dependence. In particular for nonsinglet case twist-4 
contribution is expressed as 
\begin{displaymath}  
H^{(P,D)}_{2,L}(x)=A'_2\int_x^1dzC_{2,L}(z)F^{(P,D),LT}_2(x/z,Q),
\end{displaymath}  
\begin{displaymath}  
C_2(z)=-\frac{4}{(1-z)_+}+2(2+z+6z^2)-9\delta(1-z)-\delta'(1-z),
\end{displaymath}  
\begin{displaymath}  
C_L(z)=8z^2-4\delta(1-z),
\end{displaymath}  
\begin{displaymath}  
A'_2=-\frac{2C_F}{\beta_0}\bigl[\Lambda_R\bigr]^2e^{-C},
\end{displaymath}  
where $C_F=4/3$, $C=-5/3$.
The normalization factor $\Lambda_R$ can be considered as a fitted
parameter, or, in other approach, set equal to the value of 
$\Lambda_{QCD}$. 
In Figs. 2,3 we present the IRR model predictions for $Q^2=2~GeV^2$.
At this value of $Q$ the number of active 
fermions $n_f=3$ and we set 
$\Lambda_R=\Lambda^{(3)}_{\overline{MS}}=337~MeV$, as obtained 
in our analysis.
Parameters of LT structure functions were taken as in column 2 of Table III. 
One can see that the model describes the data on $H_L(x)$ fairly well.
This in agreement with the curves, presented in Fig. 6 of Ref. \cite{IRRS} 
(remind that our value of $\Lambda$ is about 1.3 times larger, than
used in \cite{IRRS} to calculate these curves). 
At the same time there is evident discrepancy between the model
and the data on $H_2(x)$. In this connection we should note that $H_2(x)$ is 
strongly correlated with other parameters,
and in particular with $\alpha_s(M_Z)$ . In the Table V 
we present a global correlation coefficients 
$\rho_m$, calculated as  
\begin{displaymath}
\rho_m=\sqrt{1-1/e_{mm}/c_{mm}},
\end{displaymath}
where $c_{mn}$ is parameters error matrix, $e_{mn}$ -- its inverse
and indices $m,n$ runs through fitted parameters. Global
coefficients characterize the extent of correlation of a parameter
with all others parameters. (In the case of the fit with two 
parameters $\rho_1=\rho_2$ and equal to the correlation 
coefficient between these parameters). From the Table V
one can see that $H_L$ is almost uncorrelated with $\alpha_s(M_Z)$
and is correlated with other parameters less than $H_2$.
This fact can be readily understood qualitatively.
The total value of $R^{HT}$ is defined from the y-dependence of
cross sections and is weakly correlated with $\alpha_s$. 
Then, the correlation of HT contribution to $F_L$
can arise only due to  dependence of LT contribution on $\alpha_s$. 
As far, the LT contribution is small comparing with the HT one
(see Figs. 7,8) this dependence does not cause significant correlations
of the HT contribution and $\alpha_s$.
Due to lower values of $\rho$, $H_L$ is more stable in respect with 
the change of the ansatz and input of the fit.
The change of $\alpha_s$ towards higher values 
would not affect $H_L$, but can
decrease $H_2$. 

{\bf Conclusion} 

In conclusion, we performed NLO QCD fit of 
the combined SLAC/BCDMS/NMC DIS data at high $x$.
Model independent x-shape of 
high twist contribution to structure function $F_L$
is extracted. Twist-4 contribution is
found to be in fair agreement with the predictions of
infrared renormalon model. Twist-6 contribution
exhibits weak tendency to negative values, although
in the whole is compatible with zero within errors.

{\bf Acknowledgements}

I am indebted to A.L.Kataev and E.Stein for careful reading of the manuscript
and valuable comments. The work was done within the 
scientific program of the Project N99-02-16142, submitted to the 
Russian Foundation of Fundamental Research.

\end{document}